\documentclass{article}
\usepackage{psfig}
\usepackage{graphics}
\usepackage{emulateapj}
\begin{document}



\newif\ifAMStwofonts

\title{Super-Eddington Atmospheres That Don't Blow Away}
\author{Mitchell C. Begelman\altaffilmark{1,}\altaffilmark{2}}
\altaffiltext{1}{JILA, University of Colorado, Boulder, CO 80309-0440, USA; mitch@jila.colorado.edu}
\altaffiltext{2}{Also at Department of Astrophysical and Planetary Sciences, University of Colorado}
       
\date{Accepted 1999.
      Received 1999}
\begin{abstract}
We show that magnetized, radiation dominated atmospheres can support steady state patterns of density inhomogeneity that enable them to radiate at far above the Eddington limit, without suffering mass loss.  The inhomogeneities consist of periodic shock fronts bounding narrow, high-density regions, interspersed with much broader regions of low density.  The flow of radiation avoids the dense regions, which are therefore weighed down by gravity, while gas in the low-density regions is slammed upward into the shock fronts by radiation force. As the wave pattern moves through the atmosphere, each parcel of matter alternately experiences upward and downward forces, which balance on average. We calculate the density structure and phase speed of the wave pattern, and relate these to the density contrast and the factor by which the net radiation flux exceeds the Eddington limit. The presence of a magnetic field is essential for the existence of these flows, since magnetic tension shares the competing forces between regions of different densities, preventing the atmosphere from blowing apart. There appears to be a broad family of modes propagating in arbitrary directions with respect to the direction of the mean magnetic field, and exhibiting a range of density contrasts.  While the transition from low to high density occurs through a strong shock, the gas must pass through a slow magnetosonic critical point in order to return to the low-density state. 

The flux of radiation escaping from the atmosphere exceeds the Eddington limit by a factor of order the square-root of the ratio between maximum and minimum density. In principle, this factor can be as large as the ratio of magnetic pressure to mean gas pressure.  Although the magnetic pressure must be large compared to the mean gas pressure in order to support a large density contrast, it need not be large compared to the radiation pressure. These highly inhomogeneous flows could represent the nonlinear development of the ``photon bubble" instability discovered by Gammie.  If they occur in nature, these structures could have an impact on our understanding of luminous systems such as accreting compact objects and very massive stars.
  
\end{abstract}

\section{Introduction}

The Eddington limit is of fundamental importance to the study of luminous systems such as accreting compact objects, novae and gamma-ray bursts, and supermassive stars.  Hydrostatic atmospheres dominated by radiation pressure radiate at close to the Eddington limit for the central mass, provided that they do not develop density inhomogeneities on scales much smaller than the radiation-pressure scale height. Accreting black holes either swallow all the energy produced in excess of the Eddington limit (Begelman 1979), or else the pressure of escaping radiation constrains the accretion rate to an Eddington-limited value (Shakura \& Sunyaev 1973).  Including the effects of rotation leads at best to a modest enhancement in the radiation flux (Paczy\'nski \& Wiita 1980).  It is possible to exceed the Eddington limit by advecting radiation outward in a very optically thick wind, but this requires a kinetic energy flux that greatly exceeds the radiation flux (M.~Rees 1976, private communication; Meier 1982; Becker \& Begelman 1986). This is why models for 
gamma-ray bursts demand the re-conversion of kinetic energy back to radiative form via external or internal shocks, rather than relying on direct emission from the fireball itself (Rees \& M\'esz\'aros 1992, 1994; Narayan, Paczy\'nski, \& Piran 1992; Katz 1994; Sari \& Piran 1995).

It has long been known that Eddington's constraint on the maximum luminosity of flows and atmospheres can be circumvented if the density structure becomes highly inhomogeneous on small scales. While diffusive effects tend to smooth out small-scale radiation {\it pressure} gradients, the same cannot be guaranteed of the {\it density} if the gas pressure is comparatively small.  The radiation flux, which determines the force, is then inversely proportional to the local density.  Radiation flows readily through tenuous regions, while avoiding regions of high density. Depending on the distribution of high- and low-density regions, the total flux through the system can exceed the Eddington limit by a large factor (Shaviv 1998, 2000). The matter in the low-density regions would be subject to a 
super-Eddington flux, and would be blown upward.  But the flux passing through the dense regions could be sub-Eddington, resulting in a net downward force due to gravity. At worst, only the low-density component of the atmosphere --- which need not contain most of the mass 
--- would be blown away.  At best, even the low density component could be reined in, either by the inertia of the surrounding dense matter with which it interacts, or by magnetic fields coupling the high- and low-density regions. 

It has remained an open question whether low-density conduits for radiation would form spontaneously in a radiation-dominated atmosphere and, if they do, whether the coupling between low- and high-density regions would be sufficiently strong to allow the atmosphere to maintain overall dynamical equilibrium.  Prendergast \& Spiegel (1973) proposed an analogy between a radiation-supported slab and laboratory ``fluidized beds" consisting of compressed air or water forced through a bed of particles like sand. By analogy with the instabilities that occur in such flows, they (and Spiegel 1977) proposed that super-Eddington fluxes of radiation could escape from radiation-dominated slabs via ``photon bubbles," i.e., pockets of high-entropy (low-density) fluid percolating through the higher density background.  
Recent calculations have produced evidence that overstable acoustic modes may indeed exist in radiation-dominated atmospheres, although their growth rates may be rather small and their robustness uncertain (Spiegel \& Tao 1999; Shaviv 1999).  

Far more robust unstable modes appear to be present when the fluid is permeated by a moderately strong magnetic field.  The first photon bubble modes discovered (Arons 1992; Hsu, Arons, \& Klein 1997) depend on small phase shifts due to diffusion effects and are mildly overstable, although they appear to be important in regions with extremely high optical depths and strong magnetic fields, such as neutron star accretion columns (Klein et al. 1996a,b; Jernigan, Klein, \& Arons 2000). Gammie (1998) later found strongly unstable modes which appear to occur under wide-ranging conditions where radiation pressure exceeds gas pressure, such as the inner regions of accretion disks and the outer envelopes of supermassive stars. Both Arons's and Gammie's instabilities require a magnetic field strong enough to constrain the fluid motion to nearly one dimension.

In this paper we present a calculation of steady-state, inhomogeneous structure in 
radiation-dominated atmospheres. Motivated by the linear instability calculations of Gammie and Arons, we pay special attention to the effects of magnetic fields. Using a local approximation confined to scales much smaller than the scale height of the atmophere, we derive periodic, plane-wave solutions that are fully nonlinear, corresponding to trains of shock fronts sliding through the atmosphere.  Matter cycles between the dense shocked phase, which is pulled downward by gravity, and a tenuous phase which is pushed upward by radiation force.  Magnetic tension coupling the phases allows the system to maintain overall dynamical equilibrium.      

The radiation flux passing through the atmosphere is inversely proportional to the distance between shocks and exceeds the Eddington limit by a factor of order the square-root of the ratio between maximum and minimum density ($\sim$ the ratio between maximum and mean density).  This factor may be as large as $\beta^{-1} \gg 1$, the ratio between magnetic and gas pressure.  We speculate that these wave trains may represent the nonlinear evolution of Gammie's (1998) photon bubble instability.

\section{Equations and Assumptions}

The simplified equations of our model are derived from the usual MHD and radiation-hydrodynamic equations, the latter assuming the Eddington approximation with terms retained only to $O(v/c)$ in the fluid motions (Mihalas \& Mihalas 1984). Instead of modeling the radiation-gas thermal coupling (which should have little effect on the solutions we seek) in detail, we separate the gas and radiation energetics and assume an isothermal equation of state for the gas, with sound speed $c_g = (p_g / \rho)^{1/2}$. We also assume a uniform gravitational field $- g \hat z$. The basic equations are then
\begin{equation}
\label{}
{ D\rho \over Dt} = - \rho \nabla\cdot \bf v
\end{equation}
\begin{equation}
\label{mom}
\rho {D{\bf v} \over Dt} = - c_g^2 \nabla \rho + {1 \over 4\pi} (\nabla \times {\bf B}) \times {\bf B}  - \rho g \hat z + {\rho \kappa \over c} {\bf F}
\end{equation}
\begin{equation}
\label{freezing}
{\partial {\bf B} \over \partial t}  = \nabla \times (\bf v \times \bf B)
\end{equation}
\begin{equation}
\label{divb}
\nabla \cdot {\bf B} = 0
\end{equation}
\begin{equation}
\label{radflux}
{1 \over c^2} {D{\bf F} \over Dt} + {\rho \kappa \over c} {\bf F} = - \nabla p_r
\end{equation}
\begin{equation}
\label{radentropy}
3 p_r {D\over Dt} \ln \left({p_r \over \rho^{4/3}}\right) = - \nabla \cdot \bf F
\end{equation}
where $p_r$ is the radiation pressure, $\bf F$ is the radiation flux (relative to the fluid frame), and $\kappa$ is the opacity (assumed to be constant).  By neglecting the entropy of the gas in eq.~(\ref{radentropy}) we makes a fractional error of order $p_g / p_r$, which we assume to be negligible to lowest order. We neglect the $D{\bf F}/ Dt$ term in eq.~(\ref{radflux}) and simplify eq.~(\ref{radentropy}) by neglecting $|D \ln p_r / Dt|$ compared to $|D\ln \rho / Dt|$, to obtain
\begin{equation}
\label{radentropy2}
\nabla \cdot {\bf F} = - 4 p_r (\nabla\cdot {\bf v}) .
\end{equation}
The latter approximation is justified according to eq.~(\ref{radflux}) if radiation is not strongly ``trapped" by the inhomogeneities, i.e., if the condition $\tau_\lambda = \rho \kappa \lambda \ll c/v$ is satisfied, where $\lambda$ is the characteristic separation between density enhancements (the wavelength if the flow is periodic) and $\rho$ and $v$ are characteristic densities and velocities, respectively.  In this case the finite rate of radiative diffusion produces a force term that behaves like a frictional drag. 

Implicit in the idea of a radiation-supported atmosphere is an overall vertical stratification, with a pressure scale height $|\nabla \ln p |^{-1} = H \sim p/(\rho g)$. To simplify the model further, we will seek structures of sufficiently small separation that we can neglect this stratification to lowest order, i.e., we treat $\lambda/H$ as a first-order quantity.  Thus we treat $p_r$ as a constant to lowest order.  The validity of all these approximations can be checked {\it a posteriori}.  We then seek ``plane wave" solutions in which the vector quantities ($\bf v$, $\bf B$, $\bf F$) lie in the $x-z$ plane and all quantities depend on position and time through the combination
\begin{equation}
\label{sdef}
s = x + \zeta z + v_0 t. 
\end{equation}
Surfaces of constant density (i.e., constant $s$) are tilted with respect to the vertical by an angle $\theta_s = - \tan^{-1} \zeta$. We also seek solutions that are periodic in $s$ with wavelength $\lambda$.  

Using a prime to denote differentiation with respect to $s$, we write the continuity equation as
\begin{equation}
\label{cont}
v_0\rho' + (\rho v_x)' + \zeta (\rho v_z)' = 0
\end{equation}
which integrates to
\begin{equation}
\label{cont2}
v_0\rho + \rho v_x + \zeta \rho v_z = \dot M = {\rm const.}
\end{equation}
By demanding that there be no net mass flux through the atmosphere,
\begin{equation}
\label{intcond1}
\int^{s+\lambda}_s\rho v_x ds = \int^{s+\lambda}_s \rho v_z ds = 0 
\end{equation}
and defining the mean density
\begin{equation}
\label{meandens}
\int^{s+\lambda}_s\rho  ds \equiv \rho_0 \lambda,
\end{equation}
we obtain $\dot M = \rho_0 v_0$. The mean density $\rho_0$ has special significance as the value of the density for which gravity and radiation forces exactly balance.  It is also the density at which the flow must pass through a critical point (associated with the speed of slow magnetosonic waves) while moving between regions of high and low density.

The condition $\nabla \cdot {\bf B} = 0 $ is automatically satisfied if we represent ${\bf B}$ by
\begin{equation}
\label{Bdef}
B_x = B (1 + \zeta A); \ \ B_z = B(b - A)
\end{equation}
where $B$ and $b$ are constants and $A(s)$ is any periodic function of $s$ with zero mean. Using the continuity equation and demanding that $A$ integrate to zero across a wavelength, the flux-freezing equation (\ref{freezing}) integrates to
\begin{equation}
\label{Adef}
A = {\rho \over \rho_0 v_0} (b v_x - v_z). 
\end{equation}

Next we consider the radiation force term in the momentum equation. Averaging over a wavelength, the only surviving terms in the momentum equation are 
\begin{equation}
\label{Faverage}
\lambda^{-1} \int^{s+\lambda}_s \left( - \rho g \hat z + {\rho \kappa\over c} {\bf F} \right) ds = - \rho_0 g \hat z + \langle {\rho \kappa\over c} {\bf F}\rangle  = 0 ,
\end{equation}
so we write 
\begin{equation}
\label{Fdef}
{\bf F} = {c \over \kappa} {\rho_0 \over \rho} g \hat z + \zeta {c \over \kappa} {g \over \rho} (\rho - \rho_0) \hat x
+ {\bf F}_D .
\end{equation}
${\bf F}_D$, which averages to zero and has a divergence that satisfies eq.~(\ref{radentropy2}), is the portion of the flux associated with radiation drag. We obtain the remaining terms in eq.~(\ref{Fdef}) by requiring that eq.~(\ref{Faverage}) be satisfied and that the divergence of these terms vanish.  We would like the drag force to vanish when ${\bf v}\rightarrow 0 $ and to be linear in $v_x$ and $v_z$, thus guaranteeing that the average vanishes. If $p_D(s)$ is the perturbation in the radiation pressure due to drag, then eq.~(\ref{radflux}) gives ${\bf F}_D = (c / \kappa \rho) p_D' (\hat x + \zeta \hat z) $,  while eq.~(\ref{radentropy2}) can be integrated to give
\begin{equation}
\label{dragflux}
{\bf F}_D = - { 4 p_r \over 1 + \zeta^2}  (v_x + \zeta v_z) \cdot (\hat x + \zeta \hat z).
\end{equation} 
Eliminating the velocity dependence via the continuity equation, we obtain the following expression for the drag force:
\begin{equation}
\label{dragforce}
{\rho \kappa \over c} {\bf F}_D = {4 p_r \kappa v_0 \over (1 + \zeta^2) c } (\rho - \rho_0 )\cdot (\hat x + \zeta \hat z).
\end{equation}

The $x-$ and $z-$ components of the momentum equation are, respectively,
\begin{eqnarray}
\label{xmomentum}
\rho_0 v_0 v_x' + c_g^2 \rho' = {B^2 \over 4\pi} (1 + \zeta^2) A' (b-A) + \nonumber \\
\left( {4 p_r \kappa v_0 \over (1 + \zeta^2) c }  + \zeta g \right) (\rho - \rho_0 )
\end{eqnarray}
\begin{eqnarray}
\label{zmomentum}
\rho_0 v_0 v_z' + \zeta c_g^2 \rho' =  - {B^2 \over 4\pi} (1 + \zeta^2) A' (1+ \zeta A) +  \nonumber \\ 
\left({4 \zeta p_r \kappa v_0 \over (1 + \zeta^2) c } - g \right)  (\rho - \rho_0 ) .
\end{eqnarray}
From this point forward, however, it is more convenient to work in rotated coordinates oriented parallel and perpendicular to the surfaces of constant $s$ (i.e., surfaces of constant density, etc.).  We have
\begin{equation}
\label{rotv}
v_\perp = {v_x + \zeta v_z \over (1 + \zeta^2)^{1/2} } ; \ \ v_\parallel = {\zeta v_x - v_z \over (1 + \zeta^2)^{1/2} }
\end{equation}
and
\begin{equation}
\label{rotB}   
B_\perp = {B(1+ b \zeta)\over (1 + \zeta^2)^{1/2}} ; \ \ 
B_\parallel =  {B [\zeta - b + (1+\zeta^2)A] \over (1 + \zeta^2)^{1/2}}.
\end{equation}
Note that $B_\perp$ is a constant. We will also find it convenient to replace $A$ by the equivalent function
\begin{equation}
\label{phidef}
\phi \equiv {B_\parallel \over B_\perp} = {\zeta - b + (1+\zeta^2)A \over 1+ b \zeta }. 
\end{equation}
The rotated components of the momentum equation are then
\begin{equation}
\label{parmomentum}
\rho_0 v_0 v_\parallel' =  {B_\perp^2 \over 4\pi} (1 + \zeta^2)^{1/2} \phi' + (1 + \zeta^2)^{1/2} g (\rho - \rho_0 )
\end{equation}
\begin{eqnarray}
\label{perpmomentum}
\rho_0 v_0 v_\perp' + (1+\zeta^2)^{1/2} c_g^2 \rho' =  - {B_\perp^2 \over 4\pi} (1 + \zeta^2)^{1/2} \phi'\phi +  \nonumber \\ 
{4 p_r \kappa v_0 \over (1 + \zeta^2)^{1/2} c }  (\rho - \rho_0 ) .
\end{eqnarray}
The continuity equation becomes
\begin{equation}
\label{rotcont}
(1 + \zeta^2)^{1/2} \rho v_\perp = (\rho_0 - \rho) v_0,
\end{equation}
showing that $v_\perp$ vanishes where the local density equals the mean.

\subsection{Dimensionless Equations}

At this point it is convenient to non-dimensionalize the equations.  We define:
\begin{equation}
\label{nondim}
u_{\parallel,\perp} \equiv {v_{\parallel,\perp}\over v_0 }(1 + \zeta^2)^{1/2} ; \ \ \eta \equiv {\rho \over \rho_0}; \ \ y \equiv {s g \over c^2_g}; 
\end{equation}
\begin{equation}
\beta \equiv {4\pi \rho_0 c_g^2 \over B_\perp^2  }; \ \ m \equiv {v_0\over (1+ \zeta^2)^{1/2} c_g} ;  \ \  
\xi \equiv {4  p_r \kappa \over (1+\zeta^2)^{1/2} g }{c_g \over c} .\nonumber 
\end{equation}
These scalings are physically motivated. The fiducial velocity, $v_0/(1+ \zeta^2)^{1/2}$, is the pattern speed, and $m$ is the associated Mach number for the gas component.  The characteristic length scale is related to the scale height of the gas in the gravitational field. The parameter $\beta$ measures the ratio of gas pressure to magnetic pressure (smaller than the standard ``beta--parameter" of plasma physics by a factor 2), while the drag parameter $\xi$ is the ratio of radiation drag force to gravitational force (parallel to the density surfaces) for matter moving at the gas sound speed. 
The parallel and perpendicular components of the momentum equation become:
\begin{equation}
\label{parmomnodim}
m^2 u_\parallel' = {\phi' \over \beta} + (\eta - 1)
\end{equation}
\begin{equation}
\label{perpmomnodim}
m^2 u_\perp' + \eta' = - {\phi'\phi \over \beta} + m\xi (\eta - 1),
\end{equation}
where a prime now denotes differentiation with respect to $y$.  We eliminate $u_\perp$ and $u_\parallel$ by using the continuity and flux-freezing equations:
\begin{equation}
\label{contnodim}
u_\perp = {1 - \eta \over \eta}
\end{equation}
\begin{equation}
\label{freezingnodim}
u_\parallel = {\phi\over \eta} - {\zeta- b\over 1+ \zeta b} . 
\end{equation}
Differentiating and substituting into the momentum equations, we obtain
\begin{equation}
\label{parmomnodim2}
{m^2\over \eta ^2} \phi \eta' = - \phi' \left({1\over \beta} - {m^2 \over \eta}\right) - (\eta - 1)
\end{equation}
\begin{equation}
\label{perpmomnodim2}
\eta'\left( 1 - {m^2 \over \eta^2}\right) = - {\phi'\phi \over \beta} + m\xi (\eta - 1).
\end{equation}

Equations (\ref{parmomnodim2}) and (\ref{perpmomnodim2}) can be combined to yield an ordinary differential equation for $\eta'$ in terms of $\eta$ and $\phi$ plus an energy-type integral that gives an algebraic relationship between $\eta$ and $\phi$. To obtain the former we eliminate $\phi'$, yielding
\begin{eqnarray}
\label{windeq}
{\eta'\over \eta^2} \left[ m^2 \left( 1 + \phi^2 -  {\beta m^2 \over \eta} \right) - 
\eta^2 \left( 1 - {\beta m^2 \over \eta} \right)\right] =  \nonumber \\
(1 - \eta) \left[ m\xi \left( 1 - {\beta m^2 \over \eta} \right) + \phi  \right] .
\end{eqnarray}
Equation (\ref{windeq}) has the form of a ``wind equation" with a critical point at $\eta = 1$ ($\rho = \rho_0$).  In order for a solution to pass this point smoothly, the quantity in square brackets on the left-hand side must vanish simultaneously.  What is the physical significance of this critical condition? If we reintroduce dimensional variables, the critical condition can be written
\begin{equation}
\label{critcond}
\left( {v_0^2 \over 1 + \zeta^2} \right)^2 -  \left( {v_0^2 \over 1 + \zeta^2}\right) \left( {B_\perp^2 + B_\parallel^2 \over 4\pi \rho_0 } + c_g^2 \right) + {B_\perp^2 \over 4\pi \rho_0} = 0.
\end{equation}
This is just the dispersion relation for magnetosonic waves propagating perpendicular to the constant-density surfaces, where $v_0 / (1+\zeta^2)^{1/2}$ is the phase speed. But $v_0 / (1+\zeta^2)^{1/2}$ is also the speed of matter crossing the critical surface in the frame in which the density pattern is stationary (since $v_\perp = 0$ at the critical surface in the ``lab" frame; see eq.~[\ref{rotcont}]).  Thus the critical point is a true sonic point for disturbances mediated by magnetosonic waves.  Returning to dimensionless variables, the critical condition can be written
\begin{equation}
\label{critnodim}
m^2 = {1\over 2\beta} \left[ 1 + \phi_{\rm cr}^2 + \beta \pm \sqrt{(1 + \phi_{\rm cr}^2 + \beta)^2 - 4\beta }  \right],
\end{equation} 
where $\phi_{\rm cr}$ is the value of $\phi$ at the critical point and the $\pm$ sign corresponds to fast and slow magnetosonic waves, respectively. 

To derive the energy integral we eliminate $(\eta - 1)$ and integrate the resulting expression.  The result is
\begin{equation}
\label{energyint}
{\phi^2 \over 2} + m \xi \left( 1 - {\beta m^2 \over \eta} \right)\phi + 
\beta \left( \eta + {m^2 \over \eta} \right) = \varepsilon,  
\end{equation}
where $\varepsilon$ is a constant of integration. 

\section{Solutions}
\subsection{General Considerations}

We seek flows in which the density varies between high values ($\eta > 1$), where gravity dominates, and low values ($\eta < 1$) where upward radiation force dominates.  Since we also seek periodic solutions, in the stationary-pattern frame the gas must flow from high to low density and back again. In order for this to occur {\it smoothly} the flow must pass through a critical point.  It is appropriate to demand that the critical condition be satisfied for $\eta$ decreasing from high values to low values, corresponding to a smooth transition from subsonic to supersonic flow.  Since this corresponds to $d\eta / dy < 0 $ for $v_0 > 0$, we require
\begin{equation}
\label{ineqcond}
 m\xi \left( 1 - {\beta m^2 \over \eta} \right) + \phi < 0  
\end{equation} 
for the case where the magnetic field is ``strong" ($\beta m^2 < 1$) and the opposite inequality in the limit of a weak magnetic field ($\beta m^2 > 1$). Since $m$ and $\xi$ are positive by assumption, this imples that $\phi$ must be negative (positive) near the critical point, for the strongly (weakly) magnetized cases. If the flow is to be periodic in $y$, there must also be supersonic to subsonic transitions.  It is easy to show that these transitions {\it cannot} occur through a smooth crossing of the critical point, except possibly for the unphysical case $m\xi = 0$, i.e., where there is no radiation drag. According to eq.~(\ref{critnodim}), $\phi_{cr}^2$ would have to have the same value at both crossings, however, since $d\eta / dy > 0 $ for the supersonic-to-subsonic transition with $v_0 > 0$, the sign of the inequality (\ref{ineqcond}) would have to be reversed.  The only way to accomplish this is to flip the sign of $\phi_{cr}$ (i.e., of $B_\parallel$), but this is clearly incompatible with conservation of the energy integral (\ref{energyint}) if $m \xi \neq 0$.  Thus, the transition from supersonic to subsonic flow generally must occur at a shock discontinuity. 

The above argument implies that the nonlinear flows considered here cannot exist in the absence of a magnetic field. In the nonmagnetized limit ($\beta\rightarrow \infty$), eq.~(\ref{windeq}) reduces to 
\begin{equation}
\label{nonmagwindeq}
{\eta' \over \eta^2} (1 + \eta) = m\xi ,
\end{equation}
which does not permit a smooth transition from subsonic to supersonic flow (i.e., $d \eta / dy$ is positive for all $y$).  This conclusion is consistent with findings by Gammie (1998) and Arons (1992) that magnetic tension is necessary for the existence of linear ``photon bubble" instability; we shall strengthen this analogy later, in \S~4.2.  Our rejection of the nonmagnetized limit also allows us to choose a sign in the expression for the critical Mach number, eq.~(\ref{critnodim}). Since the regions of high and low density need to be coupled dynamically by magnetic tension, we presume that the speed in the pattern frame never reaches the fast magnetosonic speed and choose the negative root for slow magnetosonic waves. Adoption of the minus sign in this relation places severe constraints on the value of $\beta m^2$, the parameter that discriminates between ``strong" and ``weak" magnetic fields. First, we note that $\beta m^2$ is a monotonically decreasing function of $\phi_{\rm cr}^2$ at fixed $\beta$.  Then, setting $\phi_{\rm cr}^2 = 0$, we find that $\beta m^2 < \min (\beta, 1)$, i.e., the field is always ``strong" in this sense.  

Now we consider the shock transition. In dimensionless variables, the jump conditions across the shock are
\begin{equation}
\label{jump1}
\beta \left[ m^2 \Delta\left({1 \over \eta}\right) +  \Delta\eta\right] + {\Delta(\phi^2)\over 2}  = 0
\end{equation}
\begin{equation}
\label{jump2}
\beta m^2 \Delta\left({\phi \over \eta}\right) = \Delta \phi,
\end{equation}
where $\Delta f \equiv f_+ - f_- $ and we denote downstream (upstream) quantities by $+$ ($-$) subscripts. 
We are especially interested in the possibility of large density contrasts, $\eta_+ \gg 1 \gg \eta_- $, in which case the jump conditions may be expressed approximately as
\begin{equation}
\label{jump1approx}
\phi_+ \approx \phi_- \left( 1 - {\beta m^2 \over \eta_- } \right)
\end{equation}
\begin{equation}
\label{jump2approx}
\eta_+ \approx {m^2 \over \eta_-} \left[ 1 + \phi_-^2  \left( 1 - {\beta m^2 \over 2 \eta_- } \right) \right] .
\end{equation} 

The maximum attainable density contrast $\eta_+ / \eta_-$ is limited by the strength of the magnetic field, which must be able to resist the difference between the upward radiation force in the low-density regions and the downward gravitational force in the regions of high density. Typically the field becomes highly distorted when $\eta_- $ is smaller than $\beta m^2$, or when $\eta_+$ exceeds $\beta^{-1}$.  Thus, we expect that large density contrasts will be possible only if $\beta \ll 1$, i.e., if the magnetic pressure is much larger than the mean gas pressure.  Note that this does not mean that the magnetic pressure has to be of the same order as the radiation pressure; it may be considerably smaller.
In the limit $\beta \ll 1$ the critical condition may be approximated as
\begin{equation}
\label{critnodim2}
m^2 = {1 \over 1 + \phi_{\rm cr}^2 } + O(\beta) .
\end{equation}

To construct an explicit solution for the flow for arbitrary $\beta$ we solve eq.~(\ref{windeq}) using eq.~(\ref{energyint}) to determine $\phi (\eta)$, subject to the critical condition and two additional self-consistency conditions. First, our assumption that the function $A$ has zero mean over a wavelength $\lambda$ implies 
\begin{equation}
\label{intcond2}
\int^{\eta_+}_{\eta_-} \left( \phi - {\zeta-b\over 1 + b\zeta }\right) {d\eta \over \eta'} = 0.
\end{equation}
Second, the mean density must equal $\rho_0$, implying
\begin{equation}
\label{intcond3}
\int^{\eta_+}_{\eta_-} (1 - \eta ) {d\eta \over \eta'}  = 0.
\end{equation}
By satisfying these conditions we determine the value of $\varepsilon$ and guarantee that the shock jump conditions (\ref{jump1}) and (\ref{jump2}) are satisfied.

\subsection{Strongly Magnetized Case}

\subsubsection{Stiff-Wire Approximation}

The method of solution outlined above does not determine a unique value for $\eta_-$ (or $\eta_+$), but rather allows one to calculate $\eta_+$ in terms of $\eta_-$ (or vice versa). Thus, for specified values of $\beta$, $b$, and $\zeta$ it seems possible to find a solution for any density contrast up to the maximum value beyond which magnetic support fails. In particular, for $\beta \ll \eta_- \ll \eta_+ \ll \beta^{-1}$, the terms proportional to $\beta$ in eqs.~(\ref{windeq}) and (\ref{energyint}) are negligible throughout. The magnetic field lines are nearly straight and $\phi \approx (\zeta - b)/ (1 + b\zeta)$ is approximately constant.  Because $\phi$ must be negative according to eq.~(\ref{ineqcond}), $b-\zeta$ and $1 + b\zeta $ must have the same sign. We refer to this limit as the ``stiff-wire approximation" because the tension in the magnetic field essentially reduces the flow to one-dimensional motion.  This approximation is also relevant to the photon bubble instabilities described by Arons (1992) and Gammie (1998). 

In the stiff-wire approximation, the critical gas Mach number $m$ reduces to 
\begin{equation}
\label{critnodim3}
m \approx {|1 + b\zeta | \over (1 + b^2)^{1/2} (1 + \zeta^2)^{1/2}}
\end{equation}
and the wind equation assumes the simple form
\begin{equation}
\label{windeq2}
{\eta' \over \eta^2} (1+ \eta) = m\xi - {b - \zeta \over 1 + b\zeta } .
\end{equation}
The condition (\ref{ineqcond}) places an upper limit on the drag parameter $\xi$ consistent with a solution: 
$m\xi < (b -\zeta)/(1 + b\zeta)$, corresponding to
\begin{equation}
\label{ineqcond2}
{4p_r \kappa \over g} {c_g \over c} < {(1 + \zeta^2)(1+ b^2)^{1/2} (b- \zeta) \over (1+ b\zeta) |1 + b\zeta | } .
\end{equation}
If the drag is very strong, i.e., if $\xi \gg 1$, solutions are possible only for $b \approx - 1/\zeta$, i.e., for $|B_\parallel| \gg |B_\perp |$. The integral condition (\ref{intcond3}) is readily evaluated, yielding
\begin{equation}
\label{intcond4}
\eta_- \eta_+ = 1, 
\end{equation} 
and the wind equation is easily integrated to yield
\begin{equation}
\label{stiffint}
\left( { b - \zeta \over 1 + b\zeta } -  m\xi \right) y = {1\over \eta} - {1\over\eta_+} + 
\ln \left( {\eta_+ \over \eta }\right) ,
\end{equation} 
where $y$ is the scaled distance downstream of the shock.  The wavelength $\lambda$ (i.e., distance between shocks) is given by setting $\eta = \eta_- = 1/\eta_+$ in eq.~(\ref{stiffint}):
\begin{equation}
\label{stifflambda}
\lambda = \left( {b - \zeta \over 1 + b\zeta } -  m\xi \right)^{-1} \left( \eta_+  - {1\over \eta_+} + 
2\ln \eta_+  \right).
\end{equation}
Note that, in the submagnetosonic region between the shock and the critical surface, the density drops exponentially with $y$, as expected for an isothermal, gas pressure-dominated atmosphere in a uniform gravitational field. This is an important clue to the physical mechanism driving the modes.

\subsubsection{Weak Field Distortion}

At the next level of approximation we can calculate small distortions in the magnetic field caused by gravitational and radiation forces acting on the gas. Adopting the ordering
\begin{equation} 
\label{weakordering}
\beta \ll {\beta m^2\over \eta_- } \ll 1,
\end{equation}
we write $\phi = \phi_0 + \delta\phi$ where $\phi_0 = (\zeta - b)/(1 + b\zeta) $; and $m = m_0 + \delta m$, where $m_0^2 = (1 + \phi_0^2)^{-1} $. We linearize the energy integral (\ref{energyint}) and the critical condition (\ref{critnodim}) with $\varepsilon = \varepsilon_0 + \delta\varepsilon $, keeping terms to first order in $\beta$. We then apply the integral condition (\ref{intcond2}), using (\ref{windeq2}) for $\eta'$ and the unperturbed value of $\eta_+ = 1/\eta_-$, since the integrand is already of first order. In the integral, we keep only the leading term in powers of $1/\eta_- \gg 1$. By evaluating the perturbed critical condition (at $\eta = 1$, of course) and the energy integral at both $\eta = 1$ and $\eta = \eta_-$, we finally obtain
\begin{equation}
\label{delm}
\delta m = - \beta \left( {1 - \phi_0 m_0 \xi \over \phi_0 + m_0 \xi } \right)
{\phi_0 m_0^5 \over 2\eta_- } ,
\end{equation}
\begin{equation}
\label{delphi}
\delta\phi = {\beta \over \phi_0 + m_0 \xi }
 \left[ m_0^2 (1 - \phi_0 m_0 \xi)\left({1\over 2\eta_-} - {1\over\eta} \right) - \eta \right].
\end{equation}
It is straightforward to determine $\delta\varepsilon$ as well. 

Using eq.~(\ref{delphi}) one can verify that the jump condition (\ref{jump1approx}) is satisfied to first order. One can use the second jump condition, eq.~(\ref{jump2approx}), to determine the effect of small field distortions on the density contrast. For given $\eta_-$,  $\eta_+ = 1/\eta_- + \delta\eta_+$, where 
\begin{equation}
\label{deletaplus}
{\delta\eta_+ \over \eta_+} = - \beta \left( {4 + \phi_0^2 - 3\phi_0 m_0 \xi \over \phi_0 + m_0 \xi } \right)
{\phi_0 m_0^4 \over 2\eta_- } .
\end{equation} 
 
\subsubsection{Flows with Strong Field Distortion: Limits on Density Contrast}

The weak distortion approximation breaks down when $\delta\phi \sim \phi_0$. For arbitrary values of $\zeta$, $b \sim O(1)$ this will occur when $\eta_+ / \eta_- \sim O(\beta^{-2})$. Since $\beta m^2 / \eta_- \sim O(1)$ in this limit, one must formally retain all terms in equations (\ref{windeq}) and (\ref{energyint}) and the problem becomes analytically intractable. Note, however, that variations of the field with position are confined to the region close to the density extremes (i.e., close to the shock).  In the region $\beta \ll \eta \ll \beta^{-1}$ $\phi$ is approximately constant.  To see this we write the solution of the quadratic (\ref{energyint}) as 
\begin{eqnarray}
\label{phisoln}
\lefteqn{\phi = - m\xi\left(1- {\beta m^2\over \eta}\right) - }  \\ 
& & \sqrt{m^2\xi^2\left(1- {\beta m^2\over \eta}\right)^2 + 2\varepsilon - 2\beta \left( \eta + {m^2\over \eta} \right)} , \nonumber
\end{eqnarray}
where the minus sign is dictated by condition (\ref{ineqcond}).  The argument of the 
square-root must be positive everywhere, therefore it is insensitive to $\eta$ except (possibly) near the boundaries.

Although the magnetic field lines are straight (i.e., $\phi \approx$ const.) over much of the flow,  we may not use the approximation $\phi \approx (\zeta - b)/(1 + b\zeta)$ when $\beta m^2/\eta_- \sim O(1) $.   Instead the integral condition (\ref{intcond2}) is satisfied by large variations in $\phi (\eta)$ close to the limits of integration. To calculate these variations requires a full numerical treatment, which we reserve for a later paper.  However, one can place upper bounds on the value of $\beta m^2 / \eta_-$, and thus on the maximum density contrast, without solving for the flow explicitly. There are at least three separate constraints.

First, $\phi$ must be real throughout the flow. In order for $\phi$ to be real at $\eta_- \ll \eta \ll \eta_+$ we must have $2\varepsilon + m^2 \xi^2  > 0$. The condition for $\phi_-$ to be real is then 
\begin{equation}
\label{phirealcond}
{\beta m^2 \over \eta_-} < {1\over m^2\xi^2} \left[ 1 + m^2\xi^2  - \sqrt{1 + 2m^2\xi^2(1- \varepsilon) } \right],  
\end{equation}  
which imposes a constraint for 
\begin{equation}
\label{dconstraint}
\varepsilon < 1 + {1 \over 2 m^2\xi^2}.
\end{equation}
Second, the quantity in square brackets on the left-hand side of the wind equation (\ref{windeq}) must not change sign anywhere except at the critical point $\eta = 1$. For $\eta \ll 1$ this corresponds to 
\begin{equation}
\label{keconstraint}
{\beta m^2 \over \eta} < 1 + \phi^2 .
\end{equation}
Physically, this amounts to saying that the kinetic energy in the flow (which dominates for $\eta \ll 1$) cannot exceed the total magnetic energy.  Third, a necessary condition for the jump condition (\ref{jump2approx}) to be satisfied is 
\begin{equation}
\label{jumpsatcond}
1 + \phi_-^2 \left(1 - {\beta m^2 \over 2 \eta_- }  \right) > 0 .
\end{equation}
This condition essentially guarantees that the component of the magnetic field parallel to the shock front is not so strong that it quenches the shock. Conditions (\ref{keconstraint}) and (\ref{jumpsatcond}) can be combined into a single condition:
\begin{equation}
\label{jointconstraint}
{\beta m^2 \over \eta_- (1 + \phi_-^2) } < \min \left[  1, 2 \phi_-^{-2}  \right] .
\end{equation}

\subsection{Super-Eddington Luminosities}

Equation (\ref{stiffint}) indicates that $y$ is inversely proportional to $\eta$ for $\eta\ll 1$. Since the vertical radiation pressure gradient is uniform to lowest order, the radiation flux $\propto - \nabla p_r / \rho $ passing through the gas is also inversely proportional to the density.  Therefore, regions of different densities contribute to the total luminosity in proportion to $ y/\eta \propto \eta^{-2}$. The total flux is strongly dominated by the regions of lowest density, as pointed out by Shaviv (1998). 

To state this quantitatively, the ``mean Eddington factor" (the ratio of the spatially averaged flux to the Eddington flux $cg/\kappa$) is given by  
\begin{equation}
\label{Eddfactor}
\ell = \lambda^{-1} \int^{\eta_-}_{\eta_+}  {d\eta \over \eta \eta'} 
\end{equation}
where the wavelength (in units of $y$) is given by
\begin{equation}
\label{lambdaint}
\lambda  = \int^{\eta_-}_{\eta_+}  { d\eta \over \eta'} .
\end{equation}
To leading order in the stiff-wire approximation, the expression for the Eddington factor is remarkably simple:
\begin{equation}
\label{etaell}
\ell \approx {1  \over 2 \eta_- },
\end{equation}
independent  of the orientation of the constant-density surfaces, magnetic field direction, or strength of radiation drag.  Flows supporting large density contrasts can exceed the Eddington limit by a factor of order $1/ \eta_- \gg 1 $ without blowing away. In the strongly magnetized case this factor may be as large as $\sim \beta^{-1}$, the ratio between the magnetic pressure and the {\it mean} gas pressure.

\section{Discussion and Conclusions}

\subsection{Physical Nature of the Flows}

To understand the physical effects driving the flow, it will suffice to consider the ``stiff-wire" regime, i.e., far from maximum density contrast. To lowest order, the gas slides along straight magnetic field lines in response to the force components parallel to the field. In the subsonic (dense) region between the shock and the critical surface those forces are dominated by the actual gravity, ${\bf g} = -g \hat z $, plus a constant ``effective gravity" due to the 
density-independent portions of the radiation flux, as given by equations (\ref{Fdef}) and (\ref{dragforce}):
\begin{equation} 
\label{geff}
{\bf g}_{\rm eff} = g \left[ (\zeta + m\xi) \hat x + \zeta m \xi \hat z \right] .
\end{equation}
Projecting ${\bf g} + {\bf g}_{\rm eff}$ onto the magnetic field lines and solving the equation of hydrostatic equilibrium under the condition that $\rho$ depends on $s = x  + \zeta z$, we obtain 
\begin{equation}
\label{isotherm}
\rho \propto \exp \left[ - \left( { b - \zeta \over 1 + b\zeta } -  m\xi \right) { g s\over c_g^2}  \right],
\end{equation}
which agrees exactly with eq.~(\ref{stiffint}) in the limit $1\ll \eta \leq \eta_+$.
Thus the subsonic portion of the flow resembles an isothermal atmosphere supported partially by the magnetic field.  The density declines exponentially downstream of the shock, with an $e-$folding distance of order the ``gravitational" scale height of the gas.  

The transition to supersonic flow occurs when the exponential decline has reduced the density to $\sim \rho_0$, allowing ``radiation force" (those parts of the flux that are inversely proportional to density) to take over from ``gravity".  In the supersonic regime, $\rho \ll \rho_0$, and the density decreases more gradually, $\rho \propto 1/s$,
reaching $\rho_-$ and the next shock after traversing approximately $1/\eta_-$ gravitational scale heights.  To conserve mass flux, the speed in the supersonic region increases approximately linearly with distance.  

To further clarify the nature of the flow pattern, consider the simple case in which the density surfaces are vertical ($\zeta = 0$) and radiation drag is negligible ($\xi = 0$).  Then the subsonic flow is dominated by the real gravitational field and condition (\ref{ineqcond}) implies that $b = B_z / B_x > 0$.  Since we have assumed $v_0 > 0$ the pattern of shocks and rarefactions moves toward the left ($-x$ direction), while the magnetic field lines slope upward toward the right. Thus, the pattern of inhomogeneities runs ``downhill" along the field lines, rather than uphill, as a result of the combined action of radiation pressure and gravity. This makes physical sense: the dense regions just downstream of the shocks, which are affected more by gravity than by radiation pressure, are pulled downhill at nearly the pattern speed.  Conversely, gas in the low-density upstream regions, affected more by radiation force than gravity, are pushed uphill into the shocks by the super-Eddington flux. These features of the flow are illustrated schematically in Figure 1.  The only difference between this simple case and the more general situation with nonzero $\zeta$ and $\xi$ is that the ``gravity" force in the latter case includes a contribution from the radiation field. When this is taken into account, the material behind the shock still runs ``downhill" and intercepts material being blown upward by the radiation force, in such a way that upward and downward forces acting on any parcel of matter average to zero over time.

\begin{figure*}
\centerline{\psfig{figure=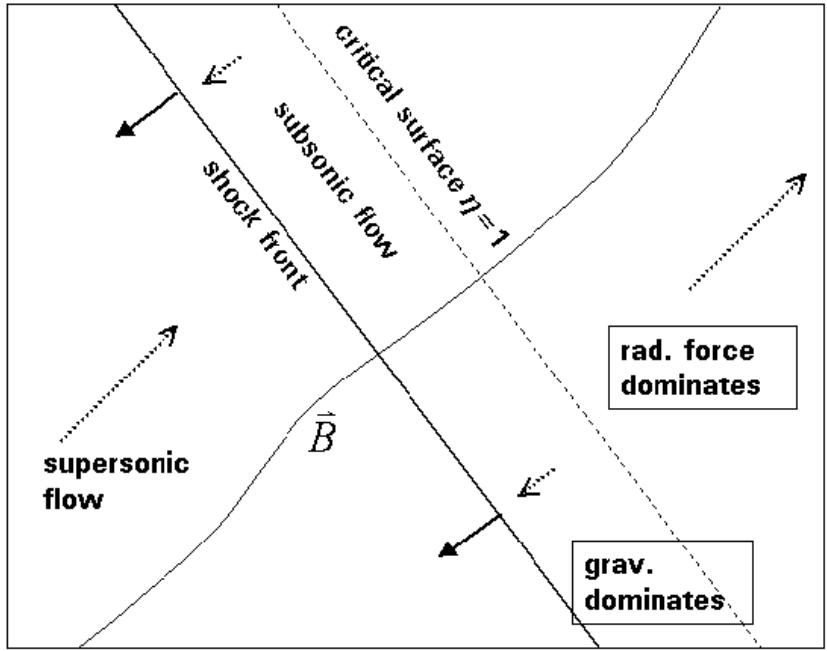,width=0.6\textwidth} }
\caption{Schematic illustration of the flows discussed in this paper. The shock front makes an angle $\tan^{-1}\zeta$ with the vertical and slides toward the left, intercepting matter blown upward and to the right by radiation force. Between the shock and critical surface matter flows downward and toward the left. Depiction of a magnetic field line illustrates the kinds of distortions caused by forces acting on the gas. }
\end{figure*}

Now let us consider weak distortions of the magnetic field. For the rest of this section we consider only the strongly magnetized limit. Equation (\ref{delphi}) combined with eq.~(\ref{ineqcond}) shows that $\delta\phi$ is positive close to the shock (i.e., where $2\eta + 2m^2/\eta > m^2/\eta_-$) and negative elsewhere.  Since $\phi \equiv B_\parallel / B_\perp$ and since $\phi_0$ is negative, this implies that the field orientation becomes more perpendicular to the constant density surfaces in the vicinity of the shock, and more parallel to the density surfaces elsewhere.  The physical significance of this result becomes clear if we revert once again to the simple case $\zeta =\xi =0$.
In this case $\phi = -b$ so a negative value of $\delta \phi$ increases the rightward slope of the field, while positive $\delta\phi$ diminishes it. Thus we see that the slope decreases on both sides of the shock. This is due to the magnetic field ``sagging" under the weight of the dense, shocked gas.  Elsewhere, the field actually steepens as the radiation force pushes the gas upwards.  The critical point lies in the region of steeper field, so it is no surprise that the critical Mach number [$\approx (1 + \phi_{\rm cr}^2 )^{-1/2} $] is smaller than it would be in the absence of field-line distortion, i.e., $\delta m < 0$ according to eq.~(\ref{delm}).

The limit of negligible radiation drag ($\xi = 0$) allows us to derive some interesting approximations that are valid for strong distortions of the magnetic field. First, we can apply the conditions (\ref{phirealcond}), (\ref{dconstraint}), and (\ref{jointconstraint}) to obtain bounds on the minimum possible density, in the form:
\begin{equation}
\label{alimits}
a_- \equiv {\beta m^2 \over \eta_-} <  \left\{ \begin{array}{ll} 
\varepsilon & 0 < \varepsilon < 1 \\
{1 + 2 \varepsilon \over 3} & 1 < \varepsilon < 4 \\
1 + {\varepsilon \over 2} - {1\over 2}\sqrt{\varepsilon(\varepsilon - 4)} & \varepsilon > 4 . \\  
\end{array} \right. 
\end{equation}
Note that $\varepsilon$ must be positive in this limit and that $a_-$ can never exceed 3, with the maximum possible value occurring for $\varepsilon = 4$.  Since $m^2 \approx (1+ 2\varepsilon)^{-1} $, the minimum possible value of $\eta_-$ continues to decrease for increasing $\varepsilon$, but as we show below (eq.~[\ref{bzero}]), in this limit the value of $\varepsilon$ is determined by the orientation of the magnetic field lines and the wavefronts.  

By expressing the (exact) critical condition in the form
\begin{equation}
\label{}
m^2(1 + 2\varepsilon) = 1 + \beta m^2 + 3\beta m^4,
\end{equation}
we can extract a factor $(1-\eta)$ from the left-hand side of the wind equation, which can be canceled with the same factor on the right-hand side to yield
\begin{equation}
\label{windeqzero}
{\eta'\over \eta^2} \left( 1 + \eta + \beta m^2 - {3\beta m^4 \over \eta} \right) = - 
  \left[ 2 \varepsilon - 2\beta \left( \eta + {m^2 \over \eta} \right) \right]^{1/2} .
\end{equation}
For large density contrasts the integrals for $\lambda$, Eddington factor $\ell$, and the normalization condition (\ref{intcond2}) relating $\varepsilon$ to $b$ are all strongly dominated by the low-density regions.  This is no surprise considering that 1) the dense slabs must be much narrower than the interslab regions, to conserve mass (since the mean density must equal $\rho_0$); and 2) most of the radiation flows through the low-density regions. Keeping only leading terms in this regime and integrating from $\eta_-$ to infinity rather than $\eta_+$, we obtain
\begin{equation}
\label{lambdazero}
\lambda \approx {1\over \beta} \left[ \sqrt{2\varepsilon} + (a_- - 1) (2\varepsilon - 2 a_-)^{1/2} \right] 
\end{equation}
where we have substituted $m^2 \approx (1 + 2\varepsilon)^{-1}$ where appropriate.  Equation (\ref{intcond2}) gives
\begin{equation}
\label{bzero}
{b - \zeta \over 1 + b \zeta }\approx { a_-(1 + 2\varepsilon) - {3\over 2} a_-^2  \over \sqrt{2\varepsilon} + (a_- - 1) (2\varepsilon - 2 a_-)^{1/2} } ,
\end{equation}
enabling one to relate the relative orientation of the magnetic field and the constant density surfaces to the ``energy parameter" $\varepsilon$.  A similar expression can be derived for the Eddington factor $\ell$.  All of these expressions reduce to the correct limits in the stiff-wire approximation, $a_- \ll 1$.

\subsection{Relationship to Gammie's Photon Bubbles}

There are tantalizing clues that the steady-state wave trains we have described may result from the nonlinear development and saturation of Gammie's (1998) photon bubble instability.  The physical mechanisms that sustain these nonlinear inhomogeneities, described in \S~4.1 above, appear to be similar to the physical effects responsible for driving Gammie's instability (see especially \S~6.3 of Gammie 1998).  In particular, Gammie's instability depends on 1) approximately one-dimensional motion enforced by the tension in the magnetic field; 2) the inverse relationship between local radiation flux and density; and 3) propagation of the pattern of density inhomogeneities {\it downhill} along field lines.  Like Gammie, we have found solutions for a wide range of wavelengths, propagation directions, and mean magnetic field orientations. Moreover, like Gammie we can show that the inhomogeneities go away in the absence of a magnetic field.

Our model does {\it not} incorporate the physics necessary to describe nonlinear developments of Arons's (1992) instability.  Arons neglects the dynamical effects of gas pressure, an approximation that amounts to taking the limit $m^2 \rightarrow \infty$ in eq.~(\ref{windeq}), with $\beta m^2$ and $m \xi$ finite.  Equations (\ref{windeq}) and (\ref{energyint}) clearly do not permit solutions that cross the critical point in this limit. The development of Arons's instability is, in fact, driven by the small but finite value of $D p_r/ Dt $, which we have neglected. 

\subsection{Validity of Approximations}

We made a number of approximations in deriving the wave structure.  First, we did not consider the thermal coupling between the gas and radiation, assuming instead an isothermal equation of state for the gas.  This approximation should be rather accurate, given that fractional perturbations in the radiation pressure (and hence in the LTE radiation temperature) are much smaller than perturbations in the gas density.  The waves are sufficiently slow that the gas should have time to come into thermal equilibrium with the radiation locally.  Such thermal effects were also found by Gammie (1998) to have negligible effect on the linear photon bubble instability.  

Second, we neglected $D{\bf F}/Dt$ and $Dp_r/Dt$ in the radiation hydrodynamic equations.  These assumptions bracket the optical depth associated with the wave motion:
\begin{equation}
\label{taucond}
{v \over c} \ll \kappa \rho \lambda \ll {c \over v} ,
\end{equation}
where $\lambda$ represents the scale length of a region with density $\rho$ and $v$ is the associated velocity relative to the pattern frame. The left-hand inequality corresponds to the neglect of $D{\bf F} / Dt$ and is most stringent in the supersonic region of the flow, where $\kappa \rho \lambda \sim  \kappa \rho_0 c_g^2 / g$ is approximately constant.  Ignoring geometric and numerical factors of order unity (and assuming $\xi \sim O(1)$), this condition becomes 
\begin{equation}
\label{taucond2}
\eta_- \gg {g \over \kappa \rho_0 c_g c } .
\end{equation}
When condition (\ref{taucond2}) is violated the diffusion of radiation may be ``flux-limited" (Mihalas \& Mihalas 1984), although the Eddington approximation (i.e., approximate isotropy of the radiation field) is still likely to be accurate. Imposing condition (\ref{taucond2}) on our flows would place a severe constraint on the maximum possible value of the Eddington factor $\ell$ (through eq.~[\ref{etaell}]) and a strong lower limit on the permissible optical depth of the atmosphere (since we seek solutions with $\eta_- \ll 1$). In fact, both of the specific cases we consider in \S~4.4 may violate this condition.  However, in practice this limit is {\it irrelevant} for the solutions we have derived. This is because $D p_r / Dt$ (i.e., radiation trapping) is negligible when condition (\ref{taucond2}) is violated, hence the radiation force term in the momentum equation (\ref{mom}) can be calculated without knowledge of the relationship between ${\bf F}$ and $\nabla p_r$.  Our flows are therefore unaffected by 
flux-limited diffusion.

The right-hand inequality of (\ref{taucond}), which corresponds to neglecting $Dp_r/Dt$ and thus radiation trapping, is most stringent in the region of exponential density gradient just behind the shock front. This condition can be written as
\begin{equation}
\label{taucond3}
{\kappa \rho_0 c_g^3 \over g c } \ll 1.
\end{equation}

Additionally, our assumption that both $p_r$ and $\nabla p_r$ are constant to lowest order amounts to sampling a region of the atmosphere much smaller than the radiation pressure scale height, $H \sim p_r/(\rho_0 g)$.  This means we can only treat wavelengths $\lambda \ll H$.  For waves with the maximum density contrast, the wavelength is of order $\beta^{-1} H_g$, where $H_g = c_g^2/g$ is the scale height associated with the gas.  Thus our models should be valid for the entire allowed range of wavelengths provided that the magnetic pressure is much larger than the mean gas pressure but smaller than the radiation pressure.  For magnetic pressure comparable to or greater than the radiation pressure, the global structure of the atmosphere would impose a serious bound on the maximum wavelengths and density contrasts.

Finally, Gammie argues that his WKB analysis is strictly valid only for wavelengths smaller than $HM_0^2$, where $M_0 \sim c/ (\kappa \rho_0 g^{1/2} H^{3/2})$ is a radiative diffusivity parameter.  In our notation this constraint corresponds to $\lambda < H_g/\xi^2 $, if we associate the wavelength of a nonlinear wavetrain with that of a linear photon bubble mode.  We have not found any special significance to this scale in the nonlinear steady-state analysis.

\subsection{Applicability to Accretion Flows}

In this section we briefly assess the applicability of our model to two important accretion problems, taking into account the effects of inhomogeneities self-consistently. 

\subsubsection{Thin Accretion Disks}

Consider first a thin, Keplerian, radiation-dominated accretion disk. For an $\alpha$-model viscosity, the dissipation rate yields a radiation flux $F \sim {3\over 2} \alpha p_r \Omega h$, where $\Omega = (GM/r^3)^{1/2}$ is the angular velocity and $h$ is the disk semi-thickness.  Assuming a vertical gravity $g \sim \Omega^2 h$ and a disk surface density $\Sigma \sim p_r /g$ we obtain the Eddington factor
\begin{equation}
\ell \sim {3\over 2} {\alpha \kappa \over c} \Sigma \Omega h .
\end{equation}   
Recall that $\ell$ is the ratio of the flux to the Eddington flux associated with the vertical component of gravity, $F_E = cg/\kappa$. Thus, a value of $\ell \gg 1$ does not necessarily imply that the disk is radiating a 
super-Eddington luminosity for the central mass, but rather that the disk is much thinner (by a factor $\sim \ell^{-1}$) than it would be if it were radiating the same flux but without inhomogeneities.  Assuming LTE, $p_r = aT^4/3$, we find that the ratio of gas pressure ($=\rho k T / \mu$) to radiation pressure is given by
\begin{equation}
{p_g \over p_r} \sim \left( {3 \over a}\right)^{1/4} {k\over \mu} \Omega^{-3/2} \Sigma^{1/4} h^{-7/4} .
\end{equation}
If the viscosity is magnetic in origin with magnetic pressure $p_M \sim \alpha p_r$, then the magnetic parameter is given by $\beta \sim p_g / p_M \sim \alpha^{-1} p_g / p_r$. This leads to an important upper limit on $\ell$, 
\begin{equation}
\label{ellthindisk}
\ell < \beta^{-1} \sim \alpha {p_r \over p_g } .
\end{equation} 
The maximum possible wavelength $\sim \alpha h$ is comfortably shorter than the scale height.

Let us now evaluate the various constraints numerically for a central mass of $m\, M_\odot$, in terms of the radius $r = x GM/c^2 = 1.5 \times 10^5 xm$. We express $\Sigma$ in units of g cm$^{-2}$, which corresponds roughly to the electron scattering optical depth. The upper limit on $\ell$ from (\ref{ellthindisk}) becomes 
\begin{equation}
\label{ellthindisk2}
\ell < 1.4 \times 10^5 \alpha \Sigma^{-1/4} \left( {h\over r}\right)^{7/4} x^{-1/2} m^{1/4}
\end{equation}
while the expression for energy generation becomes 
\begin{equation}
\label{ellthindisk3}
\ell \sim 0.6 \alpha \Sigma \left( {h\over r}\right) x^{-1/2} .
\end{equation}
Eliminating $\Sigma$ between (\ref{ellthindisk2}) and (\ref{ellthindisk3}) we obtain
\begin{equation}
\label{ellthindisk4}
\ell < 1.2 \times 10^4 \alpha  \left( {h\over r}\right)^{8/5} x^{-1/2} m^{1/5} .
\end{equation}
Note that $h/r < 1$, implying an upper limit to $\ell$.  A similar argument gives an upper limit to $\Sigma$:
\begin{equation}
\label{Sigmathindisk}
\Sigma < 1.9 \times 10^4 \left( {h\over r}\right)^{3/5} m^{1/5} \ {\rm g \ cm}^{-2}.
\end{equation}

From the self-consistency condition (\ref{taucond3}) it is straightforward to show that radiation trapping is not important for 
\begin{equation}
\label{trapthindisk}
\Sigma < 8\times 10^5 \left( {h\over r}\right)^{13/11} x^{-2/11} m^{3/11} \ {\rm g \ cm}^{-2}.
\end{equation}
By combining conditions (\ref{Sigmathindisk}) and (\ref{trapthindisk}) we can show that radiation trapping in the inhomogeneities is unimportant for
\begin{equation}
\label{trapthindisk2}
{h \over r}> 1.6 \times 10^{-3} x^{5/16} m^{-1/8} .
\end{equation}
If (\ref{trapthindisk2}) is not satisfied, then radiation trapping could be important in the densest parts of the flow, necessitating a modified treatment. 

Condition (\ref{taucond2}) becomes 
\begin{equation}
\label{fluxlimthindisk}
\ell < 5.3 \times 10^{-4} \Sigma^{9/8} \left( {h\over r}\right)^{-15/8} x^{3/4} m^{-1/8},
\end{equation}
implying that flux-limited diffusion may occur in the supersonic portions of our flows.  However, as noted earlier, this will have no effect on the dynamics of the solutions or the escaping flux.  Finally, we note that the radiation drag parameter may be expressed as
\begin{eqnarray}
\label{xithindisk}
\xi \sim 4 \kappa \Sigma \left( {c_g \over c} \right) \sim 4.3 \times 10^{-3} \Sigma^{9/8} 
\left( {h\over r}\right)^{1/8} x^{-1/4} m^{-1/8} \nonumber \\ 
< 280 \left( {h\over r}\right)^{4/5} x^{-1/4} m^{1/10},
\end{eqnarray}
where the last inequality comes from condition (\ref{Sigmathindisk}) and we have neglected the dependence on $\zeta$.

\subsubsection{Accretion onto Neutron Star Polar Cap}

We next consider supercritical accretion onto the magnetic pole of a 1 $M_\odot$ neutron star with radius $R_* = 10$ km. The surface (Newtonian) gravity is $g = 1.3 \times 10^{14}$ cm s$^{-2}$ and we assume Thomson opacity $\kappa = 0.4$ cm$^2$ g$^{-1}$, neglecting possible modifications to the opacity due to a strong magnetic field. Energy is deposited at a rate $\ell F_E = \ell cg/\kappa = 1.0 \times 10^{25}\, \ell$ erg cm$^{-2}$ s$^{-1}$, where $F_E$ is the Eddington flux.  Note that the total radiatied luminosity is then $\ell L_E (\Omega/ 4\pi)$, where $\Omega$ here is the solid angle subtended by the accretion mound. To avoid complications (such as atmospheric pressure gradients) that are outside the scope of our simple model, we consider only the region just below the accretion shock, which we assume lies $\la R_*$ above the star's surface (Arons 1992). The pressure and density are then $p_r \sim 2\ell F_E / v_{\rm ff} (R_*) = 1.2 \times 10^{15}\, \ell $ erg cm$^{-3}$ and $\rho_0 \sim 7 p_r /v_{\rm ff}^2 (R_*) = 3\times 10^{-5}\, \ell $ g cm$^{-3}$, where we have assumed a 
radiation-dominated adiabatic shock. (Note that the pressure and density may be considerably higher closer to the star.) The associated temperature and gas sound speed in LTE are $T = 2.6 \times 10^7 \, \ell^{1/4}$ K, $c_g = 6.6 \times 10^7 \, \ell^{1/8}$ cm s$^{-1}$.  The ratio of radiation pressure to gas pressure is $10^4 \, \ell^{-1/4}$, and the drag parameter is $\xi = 0.03 \, \ell^{9/8} (1 + \zeta^2)^{-1/2} $. Radiation drag becomes important only if $\ell \ga 30$, but even then it will not inhibit the development of inhomogeneities nor strongly affect the maximum luminosity.  Although our drag parameter $\xi$ may be $\ga 1$, the diffusion parameter of Arons's analysis, $M_0 \sim \xi^{-1} (p_r / p_g)^{-1/2}$, is $\ll 1$, consistent with his approximation scheme. 

For accretion onto a neutron star the magnetic field is anchored in the neutron star's crust. We expect the magnetic pressure to exceed the radiation pressure, implying that the stiff-wire approximation applies and the maximum wavelength of the inhomogeneities is limited by the scale height of the atmosphere, which we take to be $h \la R_*$.  This implies (very roughly) $\ell < 2 \times 10^3 (h/10^6 \ {\rm cm})^{4/5} $.   Trapping of radiation is not an issue here since condition (\ref{taucond3}) is satisfied for $\ell < 2.5 \times 10^4$. 

Using eq.~(\ref{etaell}) to relate $\eta_-$ to $\ell$, we find that condition (\ref{taucond2}) becomes $\ell^{1/8} \gg 11$, which is unlikely to be satisfied.  However, once again we note that flux-limited diffusion has no effect on our flow solutions when radiation trapping is unimportant. 

\subsection{Summary and Prospects for Future Work}

We have demonstrated that magnetized, radiation-dominated atmospheres can support periodic, steady-state trains of shocks with wavelengths much shorter than the radiation pressure scale height. The most important result of our analysis is that the density inhomogeneities associated with these shock patterns can allow radiation to leak out of the atmosphere at a rate that far exceeds the Eddington limit, without blowing the atmosphere apart. There are reasons to suspect that these solutions represent the nonlinear outcome of Gammie's (1998) photon bubble instability.  

The amount by which the radiation flux can exceed the Eddington limit is limited by the ratio of the magnetic pressure to the gas pressure.  It is still an open question whether dynamo action in a radiation-dominated atmosphere can amplify the magnetic pressure to a value that greatly exceeds the gas pressure (Sakimoto \& Coroniti 1989; Miller \& Stone 2000).  In neutron star accretion columns the magnetic pressure would be fixed by internal currents in the star and could greatly exceed the radiation pressure, let alone the gas pressure.  The maximum possible radiation flux might then be limited by the scale height of the atmosphere.

We have not yet shown that well-defined, persistent wavetrains can exist in real atmospheres.  Although our solutions are fully nonlinear, they are not global. The solutions formally admit arbitrary density contrasts (with corresponding wavelengths) up to a maximum imposed by the magnetic field strength, as well as nearly arbitrary directions of propagation with respect to the mean magnetic field.  Our analysis affords no indication of whether there are preferred wavevectors or wavelengths, which would be essential for establishing persistent structure.  If preferred scales and directions exist, are they set by the stability of the nonlinear solutions (about which we know nothing yet), by spatial boundary conditions associated with the finite scale height of the atmosphere, or by initial conditions associated with the development of these inhomogeneities (e.g., the fastest growing modes of Gammie's instability)? The entire class of flows may prove to be unstable, as one might worry given that they involve regions of low density accelerating into high-density slabs.  However, the fact that the interface is a shock makes it less likely that Rayleigh-Taylor instabilities would disrupt these flows.  

We hope to address these issues soon. Ideally, radiation-hydrodynamic computer simulations will be able to handle these situations within the next year or two (J. Stone 2000, private communication); Hsu et al.~(1997) have already successfully modeled the development of Arons's photon bubble instability using simplified dynamics. There is also further scope for analytic techniques to address stability by considering weak time dependence, and the effects of boundary conditions by modeling the role of the atmospheric pressure gradient more accurately.   

These questions seem well worth pursuing.  Shock trains could be important in systems as diverse as accretion disks around black holes and neutron stars, the envelopes of very massive stars, and debris from stars that have been tidally disrupted by massive black holes in galactic nuclei (Rees 1988). The generalization of these solutions to 
radiation-dominated outflows could lead to applications in novae, supernovae, and gamma-ray bursts. This phenomenon might also permit the rapid growth of supermassive black holes by accretion at high redshifts. If radiation-dominated atmospheres spontaneously develop such strong inhomogeneities that they can transmit radiation fluxes far in excess of the Eddington limit, it would have a qualitative impact on our understanding of all such systems.

\section*{Acknowledgments} 
I thank Martin Rees, Ellen Zweibel, and Gordon Ogilvie for useful discussions. This work was supported in part by NSF grants AST-9529175 and AST-9876887.

\label{lastpage}
\end{document}